%% file: main.tex
\documentclass{article}

\input{head}

\title{SPIRiT-Diffusion: SPIRiT-driven Score-Based Generative Modeling for Vessel Wall imaging}

\author{%
    Chentao Cao\thanks{Chentao Cao and Zhuo-Xu Cui contributed equally to this work.}\\
    SIAT, Chinese Academy of Sciences \\ 
    \texttt{ct.cao@siat.ac.cn} 
    \And
    Zhuo-Xu Cui\printfnsymbol{1}  \\ 
    SIAT, Chinese Academy of Sciences \\ 
    \texttt{zx.cui@siat.ac.cn} 
    \AND
    Jing Cheng  \\ 
    SIAT, Chinese Academy of Sciences \\ 
    \texttt{jing.cheng@siat.ac.cn}
    \And
    Sen Jia  \\ 
    SIAT, Chinese Academy of Sciences \\ 
    \texttt{sen.jia@siat.ac.cn}
    \And
    Hairong Zheng  \\ 
    SIAT, Chinese Academy of Sciences \\ 
    \texttt{hr.zheng@siat.ac.cn}
    \And
    Dong Liang  \\ 
    SIAT, Chinese Academy of Sciences \\ 
    \texttt{dong.liang@siat.ac.cn}
    \And
    Yanjie Zhu\thanks{Corresponding author.}  \\ 
    SIAT, Chinese Academy of Sciences \\ 
    \texttt{yj.zhu@siat.ac.cn} 
}

\begin{document}
\maketitle
\section{Synopsis}
Diffusion model is the most advanced method in image generation and has been successfully applied to MRI reconstruction. However, the existing methods do not consider the characteristics of multi-coil acquisition of MRI data. Therefore, we give a new diffusion model, called SPIRiT-Diffusion, based on the SPIRiT iterative reconstruction algorithm. Specifically, SPIRiT-Diffusion characterizes the prior distribution of coil-by-coil images by score matching and characterizes the k-space redundant prior between coils based on self-consistency. With sufficient prior constraint utilized, we achieve superior reconstruction results on the joint Intracranial and Carotid Vessel Wall imaging dataset.

\section{Introduction}

Parallel imaging reconstruction methods include sensitivity-based single combined image methods (SMASH, SENSE \cite{sodickson1997simultaneous, pruessmann1999sense} et al.) and coil-by-coil autocalibrating methods (GRAPPA, SPIRiT \cite{grappa2002, lustig2010spirit} et al.). In fact, it is very difficult to estimate the sensitivity of the coil accurately, and low-quality estimate leads to reconstruction artifacts. On the other hand, coil-by-coil autocalibrating methods utilize only the multicoil redundant prior, which leads to limited acceleration. With the proposal of score-based generative models \cite{score-based-SDE}, accurate estimation of the image prior distribution $p(\mathbf{x})$ becomes possible. However, the existing score-based methods do not consider the characteristics of multi-coil acquisition of MRI data. Therefore, we consider whether it is possible to fully exploit the information of the prior distribution of each coil while maintaining the multicoil redundant prior. We give the method that can satisfy both based on the inversion of SPIRiT. Specifically, inspired by the SPIRiT iterative reconstruction algorithm, we regard it as a reverse diffusion process, so that a forward diffusion process can be obtained, which we call SPIRiT Diffusion.

\section{Method}
MR reconstruction via the SPIRiT method can be described as the following inverse problem:
\begin{equation}
        \underset{\mathbf{x}}{\min } \frac{1}{2}\|\Phi \mathbf{\hat{x}}-\mathbf{\hat{x}}\|^{2},~~~~s.t.~~R(\mathbf{\hat{x}}) \leq \epsilon
    \label{SPIRiT}
\end{equation}
where $\mathbf{x}$ is the multicoil image domain data, $\mathbf{\hat{x}}$ is the corresponding k-space data. $\Phi$ is a series of convolution operators that convolve the entire undersampling k-space to interpolate in the missing k-space data, and $R(\mathbf{\hat{x}})$ is the data consistency constraint. Eq. \ref{SPIRiT} can be solved using the following iterative solution algorithm (iterating from $T-2$):
\begin{equation}
    \mathbf{x}_{k}=\mathbf{x}_{k+1}+ \lambda_1\mathbf{F}^{-1}((\Phi-\mathbf{I})^{H}(\Phi-\mathbf{I}) \mathbf{F}(\mathbf{x}_{k+1}))+ \lambda_2\nabla_{\mathbf{x}} R\left(\mathbf{x}_{k+1}\right) \quad k = T-2, \cdots, 0
    \label{SPIRiT iterating}
\end{equation}
Taking Eq. \ref{SPIRiT iterating} as the iteration of the reverse diffusion process, the corresponding forward diffusion process can be defined. However, this leads to the covariance of the perturbation kernel in the diffusion process cannot be calculated (the operator $\Phi$ lies in the exponential term). To solve this issue, we add a coil redundancy operator $\mathbf{Q}$ to the standard Wiener process to enforce the noises in diffusion process  satisfy the self-consistency, i.e. $\Phi(\mathbf{Q}(\mathbf{\hat{z}}))=\mathbf{Q}(\mathbf{\hat{z}})$. Here, $\mathbf{z}$ is the Gaussian noise added in the diffusion process, $\mathbf{\hat{z}}$ is the corresponding k-space. $\mathbf{Q}(\cdot)=\mathbf{S}\sum_{i=1}^n\mathbf{s}_i(\cdot)$, $\mathbf{S}$ is the $m\times n$ sensitivity matrix, $\mathbf{S}=\{\mathbf{s}_0\cdots \mathbf{s}_{n-1}\}$. The reverse diffusion process is defined as:
\begin{equation*}
    \mathrm{d} \mathbf{x}=\Big[\frac{\eta(t)}{2}\Psi\mathbf{x} -\beta(t) \mathbf{Q}^2\nabla_{\mathbf{x}} \log p_{t}(\mathbf{x})\Big]\mathrm{d} t  +\sqrt{\beta(t)}  \mathbf{Q} \mathrm{d} \mathbf{\bar w}
\end{equation*}
where $\Psi(\cdot) = \mathbf{F}^{-1}(\Phi-\mathbf{I})^{H}(\Phi-\mathbf{I})\mathbf{F}(\cdot)$, $\frac{\eta(t)}{2}\Psi(\cdot)$ is the drift coefficient of $\mathbf{x}(t)$ and $\sqrt{\beta(t)} \mathbf{Q}$ is the diffusion coefficient of $\mathbf{x}(t)$. The corresponding forward diffusion process becomes
\begin{equation*}
    \mathrm{d} \mathbf{x}=\frac{\eta(t)}{2}\Psi \mathbf{x} \mathrm{d} t+\sqrt{\beta(t)} \mathbf{Q} \mathrm{d} \mathbf{w}
\end{equation*}
The perturbation kernel of SPIRiT-Diffusion can be derived as
\begin{equation}
    p_{0 t}(\mathbf{x}(t) \mid \mathbf{x}(0)) = \mathcal{N}\left(\mathbf{x}(t) ; \mathbf{x}(0),\frac{1}{2} \int_{0}^{\tau} \beta(\tau)e^{\int_{\tau}^{t} \eta(s) d s} d \tau \mathbf{Q}^2\right)
    \label{perturbation kernel}
\end{equation}
With Eq. \ref{perturbation kernel}, the score model can be trained via
\begin{equation*}
    \boldsymbol{\theta}^*=\underset{\boldsymbol{\theta}}{\arg \min } \mathbb{E}_t\left\{\lambda(t) \mathbb{E}_{\mathbf{x}(0)} \mathbb{E}_{\mathbf{x}(t) \mid \mathbf{x}(0)}\left[\left\|\mathbf{s}_{\boldsymbol{\theta}}(\mathbf{x}(t), t)-\nabla_{\mathbf{x}(t)} \log p_{0 t}(\mathbf{x}(t) \mid \mathbf{x}(0))\right\|_2^2\right]\right\}
\end{equation*}


\section{Results}
\begin{figure}
    \centering
    \includegraphics[width=\textwidth]{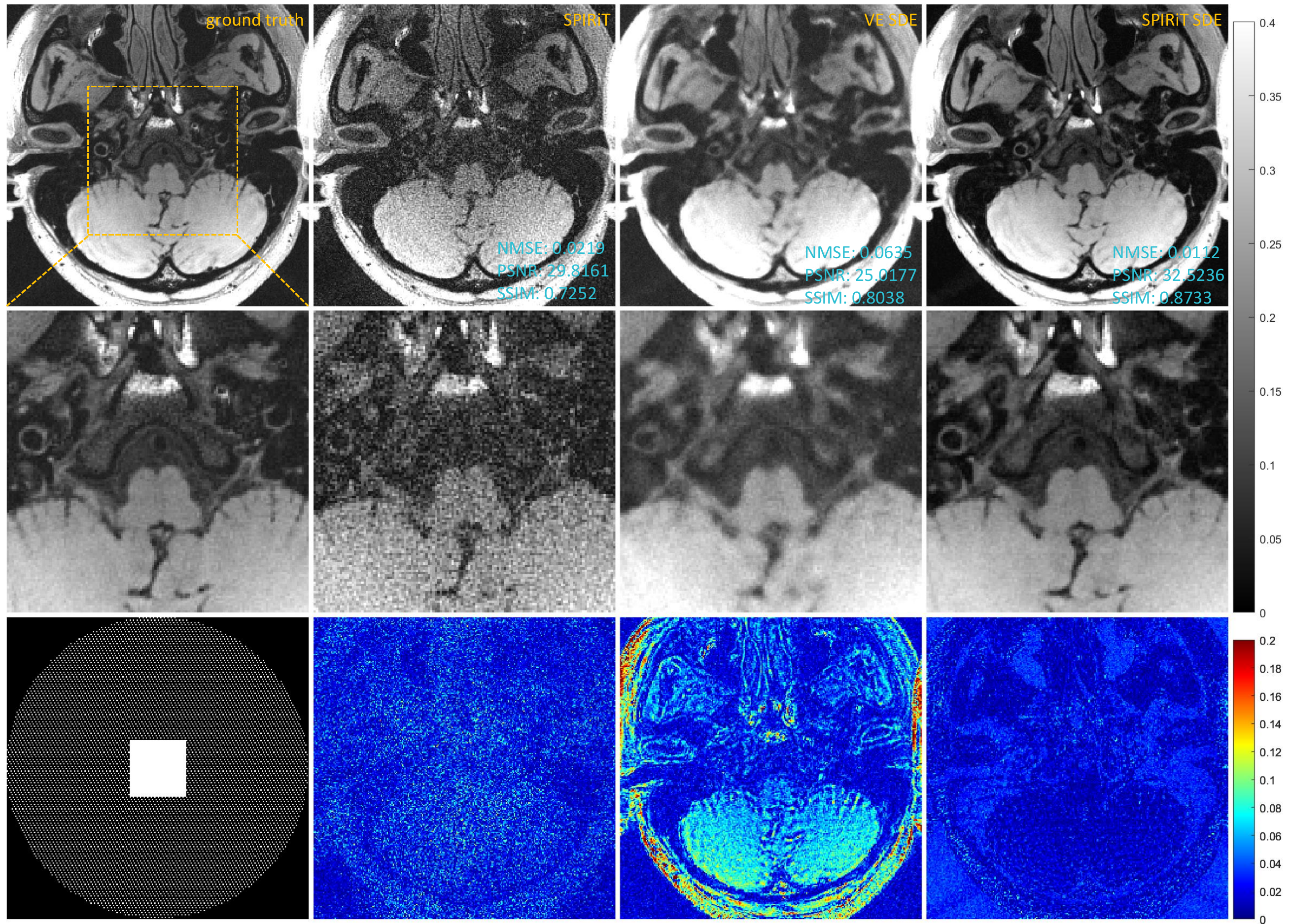}
    \caption{The reconstruction results of Vessel Wall imaging data at uniform undersampling of 6-fold.}
    \label{6-fold}
\end{figure}

\begin{figure}
    \centering
    \includegraphics[width=\textwidth]{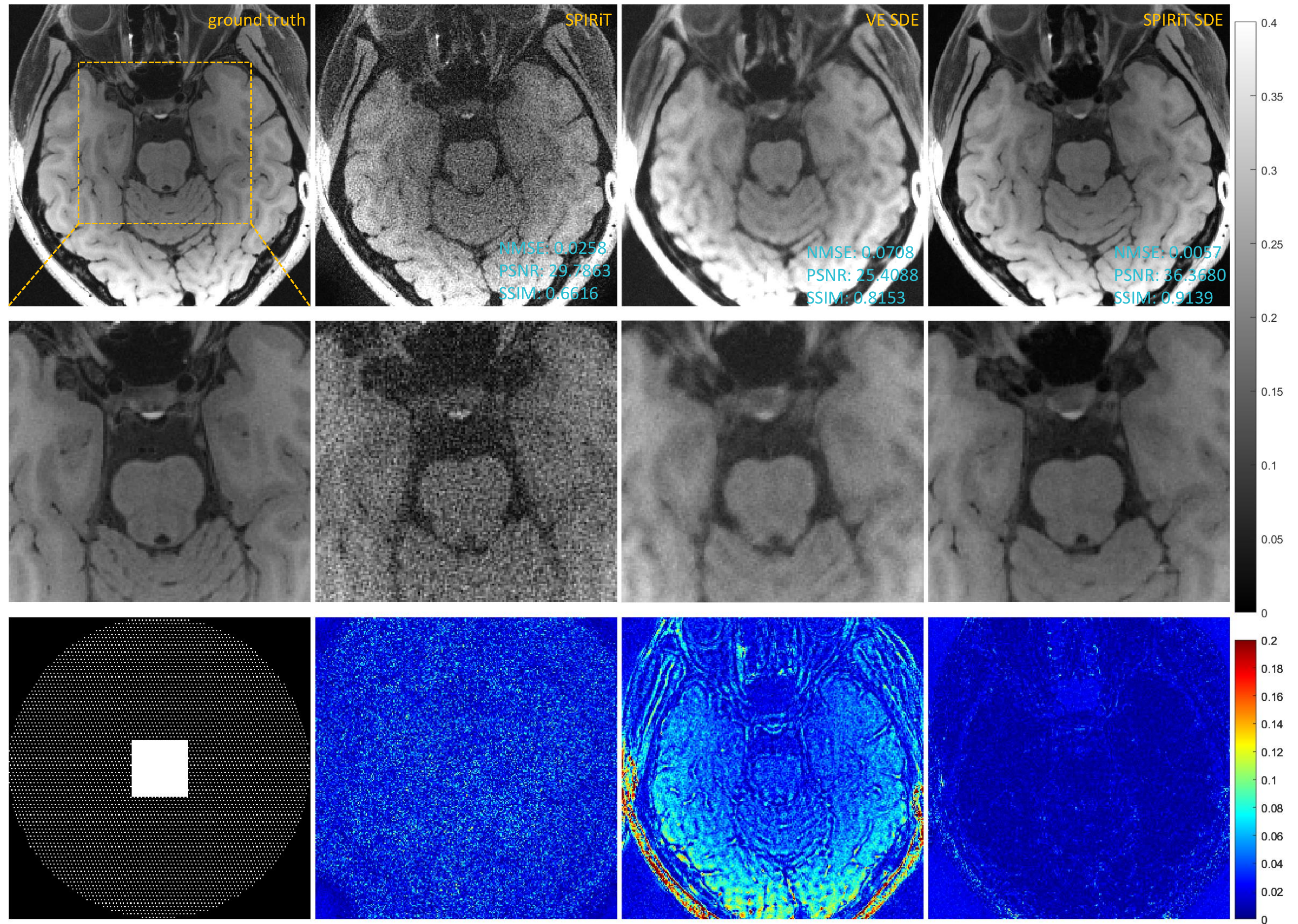}
    \caption{The reconstruction results of Vessel Wall imaging data at uniform undersampling of 10-fold.}
    \label{10-fold}
\end{figure}

We conducted experiments on 3D joint Intracranial and Carotid Vessel Wall imaging (VWI) data, which were collected on a 3T scanner (uMR 790, United Imaging Healthcare, China). The results reconstructed using SPIRiT and VE-SDE (coil-by-coil reconstruction) were shown for comparison. Fig. \ref{6-fold} shows the reconstruction results of 6-fold undersampling of the intracranial structure. SPIRiT-Diffusion reconstructs the vessel wall details well. The results of 2D 10-fold  undersampling are shown in Fig. \ref{10-fold}. Even under such extreme undersampling conditions, SPIRiT-Diffusion can recover the details of the Vessel Wall. Noted that the noise in the ground truth is serious, and SPIRiT-Diffusion also acts as a denoising agent, so SPIRiT-Diffusion shows larger errors than VE-SDE in the part of the error map where there is no anatomical structure.

\section{Conclusions}
The proposed SPIRiT-Diffusion achieved superior performance in joint Intracranial and Carotid Vessel Wall imaging, and the experiments show that our method outperforms SPIRiT and VE-SDE.
\section{Acknowledgments}
This study is supported by the National Key R\&D Program of China no. 2020YFA0712200, National Natural Science Foundation of China under grant no. 81971611, 62125111, 81901736, 81830056, 61671441, 81971611, 12026603, 62106252, 62206273 and U1805261; Shenzhen Science and Technology Program under grant no. RCYX20210609104444089.

\clearpage
\bibliography{refs}

\end{document}

%% file: head.tex
\usepackage[preprint]{neurips_2022}

\usepackage[utf8]{inputenc} 
\usepackage[T1]{fontenc}    
\usepackage{hyperref}       
\usepackage{url}            
\usepackage{booktabs}       
\usepackage{amsfonts}       
\usepackage{nicefrac}       
\usepackage{microtype}      
\usepackage{xcolor}         

\usepackage{enumitem}
\makeatletter
\newcommand{\printfnsymbol}[1]{%
  \textsuperscript{\@fnsymbol{#1}}%
}
\usepackage{algorithmicx,algpseudocode,algorithm}
\usepackage{wrapfig,tikz}
\usepackage{amsmath,longtable,fancyhdr,booktabs,multirow,graphicx,float}
\usepackage{adjustbox, bigstrut, tabularx, multirow, makecell, diagbox}
\usepackage{amssymb,xcolor,amsthm}
\usepackage{subcaption}

\newcommand{\be}{\begin{equation}}
	\newcommand{\ee}{\end{equation}}

\definecolor{Gray}{gray}{0.85}
\definecolor{LightCyan}{rgb}{0.88,1,1}

\newcolumntype{a}{>{\columncolor{Gray}}c}

\makeatletter
\usepackage{xspace}
\def\@onedot{\ifx\@let@token.\else.\null\fi\xspace}
\DeclareRobustCommand\onedot{\futurelet\@let@token\@onedot}